\documentclass[conference]{IEEEtran}
\IEEEoverridecommandlockouts
\usepackage{cite}
\usepackage{amsmath,amssymb,amsfonts}
\usepackage{graphicx}
\usepackage{textcomp}
\usepackage{xcolor}

\usepackage{todonotes}
\usepackage[hyphens]{url} 
\usepackage{hyperref} 
\usepackage{booktabs} 
\usepackage{caption} 
\usepackage{subcaption} 
\usepackage{algorithm}
\usepackage{algpseudocode}
\usepackage{numprint}

\usepackage{listings, xcolor}

\definecolor{verylightgray}{rgb}{.97,.97,.97}

\lstdefinelanguage{Solidity}{
	keywords=[1]{anonymous, assembly, assert, balance, break, call, callcode, case, catch, class, constant, continue, constructor, contract, debugger, default, delegatecall, delete, do, else, emit, event, experimental, export, external, false, finally, for, function, gas, if, implements, import, in, indexed, instanceof, interface, internal, is, length, library, log0, log1, log2, log3, log4, memory, modifier, new, payable, pragma, private, protected, public, pure, virtual, push, require, return, returns, revert, selfdestruct, send, solidity, storage, struct, suicide, super, switch, then, this, throw, transfer, true, try, typeof, using, view, while, with, addmod, ecrecover, keccak256, mulmod, ripemd160, sha256, sha3}, 
	keywordstyle=[1]\color{blue}\bfseries,
	keywords=[2]{address, bool, byte, bytes, bytes1, bytes2, bytes3, bytes4, bytes5, bytes6, bytes7, bytes8, bytes9, bytes10, bytes11, bytes12, bytes13, bytes14, bytes15, bytes16, bytes17, bytes18, bytes19, bytes20, bytes21, bytes22, bytes23, bytes24, bytes25, bytes26, bytes27, bytes28, bytes29, bytes30, bytes31, bytes32, enum, int, int8, int16, int24, int32, int40, int48, int56, int64, int72, int80, int88, int96, int104, int112, int120, int128, int136, int144, int152, int160, int168, int176, int184, int192, int200, int208, int216, int224, int232, int240, int248, int256, mapping, string, uint, uint8, uint16, uint24, uint32, uint40, uint48, uint56, uint64, uint72, uint80, uint88, uint96, uint104, uint112, uint120, uint128, uint136, uint144, uint152, uint160, uint168, uint176, uint184, uint192, uint200, uint208, uint216, uint224, uint232, uint240, uint248, uint256, var, void, ether, finney, szabo, wei, days, hours, minutes, seconds, weeks, years},	
	keywordstyle=[2]\color{teal}\bfseries,
	keywords=[3]{block, blockhash, coinbase, difficulty, gaslimit, number, timestamp, msg, data, gas, sender, sig, value, now, tx, gasprice, origin},	
	keywordstyle=[3]\color{violet}\bfseries,
	identifierstyle=\color{black},
	sensitive=false,
	comment=[l]{//},
	morecomment=[s]{/*}{*/},
	commentstyle=\color{gray}\ttfamily,
	stringstyle=\color{red}\ttfamily,
	morestring=[b]',
	morestring=[b]"
}

\lstdefinestyle{solidity-style}{
	language=Solidity,
	backgroundcolor=\color{verylightgray},
	extendedchars=true,
	basicstyle=\footnotesize\ttfamily,
	showstringspaces=false,
	showspaces=false,
	numbers = none,
	tabsize=2,
	breaklines=true,
	showtabs=false,
	captionpos=b
} 
\lstdefinestyle{text-style}{
    basicstyle=\footnotesize\ttfamily,
	backgroundcolor=\color{verylightgray},
	extendedchars=true,
	showstringspaces=false,
	showspaces=false,
	numbers=none,
	tabsize=2,
	showtabs=false,
	captionpos=b,
	breaklines=true,
}

\newcommand{\ethersolve}{EtherSolve}
\newcommand{\pushed}{pushed}

\newcommand{\todonew}[1]{#1}
\usepackage{todonotes}

\def\BibTeX{{\rm B\kern-.05em{\sc i\kern-.025em b}\kern-.08em
    T\kern-.1667em\lower.7ex\hbox{E}\kern-.125emX}}
    
\begin{document}

\title{\ethersolve{}: Computing an Accurate Control-Flow Graph from Ethereum Bytecode}

\author{\IEEEauthorblockN{Filippo Contro}
\IEEEauthorblockA{\textit{Computer Science department} \\
\textit{University of Verona}\\
\small{filippo.contro\_01@studenti.univr.it}}
\and
\IEEEauthorblockN{Marco Crosara}
\IEEEauthorblockA{\textit{Computer Science department} \\
\textit{University of Verona}\\
\small{marco.crosara@studenti.univr.it}}
\and
\IEEEauthorblockN{Mariano Ceccato}
\IEEEauthorblockA{\textit{Computer Science department} \\
\textit{University of Verona}\\
mariano.ceccato@univr.it\\
{\small ORCID: 0000-0001-7325-0316}}
\and
\IEEEauthorblockN{Mila Dalla Preda}
\IEEEauthorblockA{\textit{Computer Science department} \\
\textit{University of Verona}\\
mila.dallapreda@univr.it\\
{\small ORCID: 0000-0003-2761-4347}}
}


\maketitle

\begin{abstract}
Motivated by the immutable nature of Ethereum smart contracts and of their transactions, quite many approaches have been proposed to detect defects and security problems before smart contracts become persistent in the blockchain and they are granted control on substantial financial value. 

Because smart contracts source code might not be available, static analysis approaches mostly face the challenge of analysing compiled Ethereum bytecode, that is available directly from the official blockchain. However, due to the intrinsic complexity of Ethereum bytecode (especially in jump resolution), static analysis encounters significant obstacles that reduce the accuracy of exiting automated tools.

This paper presents a novel static analysis algorithm based on the symbolic execution of the Ethereum operand stack that allows us to resolve jumps in Ethereum bytecode and to construct an accurate control-flow graph (CFG) of the compiled smart contracts. \ethersolve{} is a prototype implementation of our approach. Experimental results on a significant set of real world Ethereum smart contracts show that \ethersolve{} improves the accuracy of the execrated CFGs with respect to the state of the art available approaches. 

Many static analysis techniques are based on the CFG representation of the code and would therefore benefit from the accurate extraction of the CFG. \todonew{For example, we implemented a simple extension of \ethersolve{} that allows to detect instances of the re-entrancy vulnerability.}

\end{abstract}


\section{Introduction}
\label{sec:intro}

Smart contracts are a recent extension to cryptocurrencies (e.g., Ethereum) that allows programs to be stored in the blockchain and to be executed by the distributed network of miners. Thus, a smart contract is a self-executing program that runs on a blockchain~\cite{book-Introducing-Ethereum}.

The most peculiar feature of a smart contract is that the program and all its transactions are immutable, i.e., once written to the distributed blockchain, they cannot be updated, even in case of programming defects identified after deployment~\cite{sai2020inheritance}. Programming defects and vulnerabilities might result in frauds or in financial values of cryptocurrencies that are frozen forever~\cite{dao1}.
Hence, code review, possibly supported by automated analysis tool, is crucial to detect programming errors and vulnerabilities before defective smart contracts are used or erroneous transactions are committed and permanently stored in the blockchain. This is especially critical for closed source smart contracts, whose source code can not be inspected by a client, and the only available representation is the compiled bytecode~\cite{li2020stan}.



The accuracy of static analysis is one of the key points to promptly deploy and run correct smart contracts. However, tools from the state of the art for detecting programming defects and vulnerabilities in Ethereum smart contracts experience substantial limitations, and their results contain worrying amount of false positives (false alarms) and false negatives (overlooked problems)~\cite{ghaleb2020effective}. A potential explanation for such poor performance could be  the intrinsic difficulty of analysing Ethereum bytecode. In fact, despite the bytecode is easy to parse (fixed length opcodes)~\cite{yellow-paper} its semantics and control-flow graph (CFG) are difficult to reconstruct due to specific language design choices:
\begin{itemize}
\item Jump destination is not an opcode parameter, but a jump opcode assumes the destination address to be available on the stack, dynamically computed by previous code;
\item There is no opcode for returning from functions: {\em return} is implemented by pushing the return address to the stack, and then performing a jump;
\item Functions are removed by the compiler. Intra-contract function calls are replaced by jumps. Inter-contract function calls are resolved by a {\em dispatcher} at the smart contract entry point, that decides what address to jump to, depending on the call's actual parameters;
\item The smart contract constructor is executed only once, when the contract is initially deployed in the blockchain and then discarded. Thus, the constructor bytecode is not available in the blockchain~\cite{soliditydoc-creatingcontracts}.
\end{itemize}
The accurate extraction of the CFG from the bytecode is at the basis of the success of many static analysis algorithms. 
In this paper we focus on the extraction of the CFG from the Ethereum bytecode. To this end, we propose a static analysis approach, called {\em symbolic stack execution}, that resolves jump destinations based on the symbolic execution of the operand stack. After jump destinations are resolved, an accurate CFG can be built. This approach has been implemented in  \ethersolve{}\footnote{The tool is available on github.com/SeUniVr/EtherSolve/tree/ICPC-2021}~\cite{ethersolve,replication-package}: a fully automated tool that provides a beneficial starting point for any sophisticated static analysis meant to identify programming defects or vulnerabilities.
\todonew{Indeed, starting from the results of \ethersolve{}, we implemented a component to detect occurrences of re-entrancy vulnerability in Ethereum smart contracts. This detector turns out to be comparable or superior to state-of-the-art security scanning tools.}

The paper is structured as follows. After covering the  background of smart contracts in Section~\ref{sec:background}, the different phases of our analysis are described in Section~\ref{sec:analysis}. Section~\ref{sec:experiment} presents our empirical validation and comparison with state of the art tools. Then, Section~\ref{sec:related} discusses related work and Section~\ref{sec:conclusion} closes the paper.
\section{Background}
\label{sec:background}



Ethereum is a global, open-source platform for decentralised applications, based on a blockchain technology. On Ethereum network, it is possible to write simple programs, called \emph{smart contracts}~\cite{book-Introducing-Ethereum},\cite{soliditydoc},\cite{Zheng_2020}, that control the cryptocurrency called Ether (ETH)~\cite{ethereum-site}.
The actions that can be performed in Ethereum are basically transactions, i.e. movements of funds or data between different accounts. Every new transaction is irreversible and it is permanently added in a new \emph{block} that updates the blockchain~\cite{book-Introducing-Ethereum},\cite{book-mastering-ethereum}.




\subsection{Solidity Language}

\begin{lstlisting}[style=solidity-style,caption={Solidity code example},label={lst:solidity}]
pragma solidity ^0.6.0;
contract SimpleBank {
    mapping(address => uint256) private balances;
    function deposit(uint256 amount) public payable{
        require(msg.value == amount);
        balances[msg.sender] += amount;
    }
    function deposit100() public payable{
        require(msg.value == 100);
        balances[msg.sender] += 100;
    }
    function withdraw(uint256 amount) public{
        require(amount <= balances[msg.sender]);
        balances[msg.sender] -= amount;
        msg.sender.transfer(amount);
    }
}
\end{lstlisting}

Solidity is the most widely spread programming language for Ethereum: it is an object-oriented, high-level and Turing-complete language for implementing smart contracts~\cite{soliditydoc}, \cite{etherscan}.

Solidity contracts are basically objects with functions and fields. The example in Listing~\ref{lst:solidity} reports a smart contract written in Solidity to implement a bank.
The field {\em balances} stores the internal state of the smart contract. It is a key-value map that associates every address to an integer value that represents the funds own by the address. The functions \textit{deposit} and \textit{deposit100} allow the user to deposit currency into its virtual account. The former allows to deposit an arbitrary amount, the latter is a special case which allows to transfer exactly 100 Wei ($10^{-16}$ Ether).
The \textit{withdraw} function allows the user to get back a certain amount of Ether previously deposited. Solidity provides different primitives to interact with the blockchain environment: \emph{transfer} sends Ether to a certain address, \emph{revert} makes the transaction fail and roll back to the state prior to the transaction, \emph{require} enforces a certain boolean condition and in case the condition is not met it performs a {\em revert}, and many others not shown in the example.

\vspace{0.2cm} 
\subsection{Compiling Solidity into Ethereum Bytecode}

Before running a smart contract in the Ethereum blockchain, Solidity source code needs to be compiled to Ethereum bytecode to be executed by the Ethereum Virtual Machine (EVM).

\textbf{Constructor.}
The Solidity compiler, namely \textit{solc}, generates the \emph{creation code}. This is the constructor of the smart contract that performs the initial operations and deploys the \emph{runtime code} on the blockchain; the constructor code is then discarded and not stored in the blockchain~\cite{soliditydoc-creatingcontracts},\cite{ethervm}. 
%

\textbf{Runtime Code.}
The runtime code can be divided into three main segments.
The first segment contains the opcodes that the EVM executes; the second one is optional and contains static data (e.g., strings or constant arrays); the last segment contains the metadata, such as compiler version and different hashes of the code. The structure of metadata has been continuously changing through the different versions of the Solidity compiler.


{\bf Application Binary Interface.}
The Solidity compiler emits also the Application Binary Interface (ABI). This file contains the list of the functions in the smart contracts that can be called by a user, together with type and number of parameters. Functions are not identified by their name but by the hash of the signature. The ABI file is not deployed in the blockchain and it needs to be distributed separately as it contains the main information for interacting with a smart contract.

{\bf Stack.}
The stack is the main memory of a smart contract, it is a volatile LIFO queue with 1024 blocks of 32 bytes~\cite{soliditydoc},\cite{ethervm}. The execution relies heavily on it, as arithmetic and logic operations follow the reverse polish notation, where the data are loaded into the stack before the operation~\cite{medium-evm-runtime-env}. For instance, the bytecode \texttt{6005600301} is parsed into \texttt{PUSH1 0x05 PUSH1 0x03 ADD}, and the EVM execution will (i) push a byte to the stack containing the value \texttt{0x05}; (ii) push the value \texttt{0x03} and then (iii) execute the addition operation, which consumes two elements from the stack and leaves their sum as result, leaving the final stack with only the value \texttt{0x08}.

{\bf Opcodes.}
The complete list of opcodes with their semantics is defined in Ethereum's yellow paper~\cite{yellow-paper}, and there can be little variations among different EVM versions.

\begin{lstlisting}[style=text-style,caption={Bytecode example},label={lst:bytecode}]
Runtime Code:
6080604052600436106100345760003560e01c8063140e9ac714
610039 ... 600020600082825401925050819055505056fe
Metadata:
a2646970667358221220e62b6e0d256ecbc0a1b39b99bf0a2b50
9ed60dd83c71541b2d00fed1bde5a9e464736f6c634300060b00
33
\end{lstlisting}

\begin{lstlisting}[style=text-style,caption={Opcodes example},label={lst:opcodes}]
PUSH1 0x80 PUSH1 0x40 MSTORE PUSH1 0x4 CALLDATASIZE 
LT PUSH2 0x34 JUMPI PUSH1 0x0 CALLDATALOAD PUSH1 
0xE0 SHR DUP1 PUSH4 0x140E9AC7 EQ PUSH2 0x39 JUMPI 
... 
PUSH1 0x0 KECCAK256 PUSH1 0x0 DUP3 DUP3 SLOAD ADD 
SWAP3 POP POP DUP2 SWAP1 SSTORE POP POP JUMP INVALID
\end{lstlisting}

Listing \ref{lst:bytecode} shows a portion of the Solidity smart contract compiled into bytecode, while Listing \ref{lst:opcodes} shows the translation of the bytes into EVM opcodes. 
Ethereum bytecode can be easily parsed into opcodes, which are the minimum instructions that the EVM can execute and are identified with a byte. 

Every opcode pushes or pops a certain number of elements from/to the stack, and it can either access memory, get information about the execution environment or interact with other blockchain smart contracts.
The only opcodes with a parameter are those in the \texttt{PUSH} family: the value that the EVM pushes into the stack is taken directly from the bytes following the opcode. There are different variants of \texttt{PUSH}, depending on the number of bytes that needs to be pushed to the stack, varying from \texttt{PUSH1} (1 byte is pushed) to \texttt{PUSH32} (32 bytes are pushed)~\cite{yellow-paper},\cite{ethervm}.

A portion of the code can be used as read-only data; in fact with the \texttt{CODECOPY} opcode the execution can copy a portion of the code to the memory and then treat it as data.~\cite{vandal2}. Thus, parsing this segment of memory as code might generate spurious results, including invalid opcodes and wrong jump destinations.

{\bf Control flow opcodes.}
The control flow of the program is managed through the stack. In fact, in order to jump among different portions of the code both the \texttt{JUMP} and the \texttt{JUMPI} opcodes (unconditional and conditional jumps) read the jump destination from the stack. These destinations are not managed with labels, but with the offset of the next instruction in the code~\cite{vandal2}.
Unlike x86 assembly, in the EVM there is no concept of \emph{function}: everything is managed through jumps, and there are neither opcodes for function call nor for return from function call. The only return available is for function calls coming from external smart contracts.

These design decisions make the EVM bytecode difficult to analyse statically. In particular, since jump destination are computed at run time, the CFG cannot be reconstructed without a sort of stack simulation whose precision directly affects the accuracy of extracted CFG.

{\bf Dispatcher.}
When a transaction starts the execution of a smart contract, it can send both funds and information as \emph{call data}. 
In order to transfer the control to the code corresponding to the intended function, the compiler adds a \emph{dispatcher} at the beginning of the contract code. 
When a caller is willing to execute a certain contract function, it sends a transaction that contains the hash of the function signature, so that the dispatcher can compare it with all the hashes of the smart contract functions and then take the execution to the begin of the corresponding function code. Instead, if no call data are supplied or none of the hashes matches, the dispatcher takes the execution to the beginning of the \emph{fallback function}. This function has no parameter and returns no value~\cite{soliditydoc}.

Every function call in the Solidity contract is translated by the compiler into a sequence of \texttt{PUSH} opcodes, followed by a \texttt{JUMP} and a \texttt{JUMPDEST}. This sequence loads into the stack the return address of the calling context, the (optional) actual parameters and the address of the function to call. Then \texttt{JUMP} executes the function body, that eventually consumes the parameters from the stack, leaving the return address which, once executed, brings the execution to the \texttt{JUMPDEST}, resulting in an actual return statement. In the following sections we provide a simple example of this pattern.


\todonew{\subsection{Re-entrancy Vulnerability}}

\todonew{\textit{Re-entrancy} is one of the most prominent vulnerabilities in Solidity, because it was exploited by the infamous DAO incident~\cite{dao1} that caused serious consequences to the whole Ethereum network. This vulnerability consists in re-entering a paying function multiple times while the contract is in an inconsistent state, thus causing possible leak of funds~\cite{ghaleb2020effective}. This vulnerability might be present when a contract state update follows (instead of preceding) a fund send primitive (i.e., a {\tt call}).}
\todonew{An example of vulnerable contract is shown in Listing~\ref{lst:re-entrancy}, where the \texttt{call} statement at line~7 precedes the update of variable \texttt{bank} at line~8.}

\begin{lstlisting}[style=solidity-style, numbers=left, numberstyle=\footnotesize, numbersep=-10pt, caption={Re-entrancy example},label={lst:re-entrancy}]
   contract Bank {
       // Mapping of money owned by an address
       mapping (address => uint) bank; 
       // Function to withdraw money
       function withdraw (uint amount) public {
           require(amount <= bank[msg.sender]);
           if (msg.sender.call.value(amount)(""))
               bank[msg.sender] -= amount;
       }
   }
\end{lstlisting}

\section{Static Analysis}
\label{sec:analysis}

\ethersolve{}~\cite{ethersolve} aims at extracting the CFG from the bytecode of Ethereum smart contracts.
The CFG is a directed graph representing the flow of execution:  nodes are the program's basic blocks (sequence of opcodes with no jumps, other than in the last opcode) and edges connect potential successive basic blocks.  


\subsection{Approach Overview}

The static analysis implemented in our approach is composed of the following steps: 
\begin{itemize}
\item {\em Bytecode parsing:} The binary representation of the Ethereum bytecode is split between code and metadata; the code is then parsed to identify opcodes;
\item {\em Basic blocks identification:}  Opcodes are grouped in basic blocks and the easier jumps between basic blocks are resolved;
\item {\em Symbolic stack execution:} The execution stack is subject to symbolic execution in order to resolve the more difficult jump destinations;
\item {\em Static data separation:} The static data segment is separated from the actual executable code;
\item {\em CFG decoration}: The so obtained CFG is decorated to highlight the dispatcher and to identify the entry point of the fallback function. 
\end{itemize}
%
In the following, we describe these steps in details referring to the \emph{deposit100} function of the \emph{SimpleBank} smart contract in Listing \ref{lst:solidity}.


\begin{figure}[btp]
    \centerline{\includegraphics[width=0.9\columnwidth]{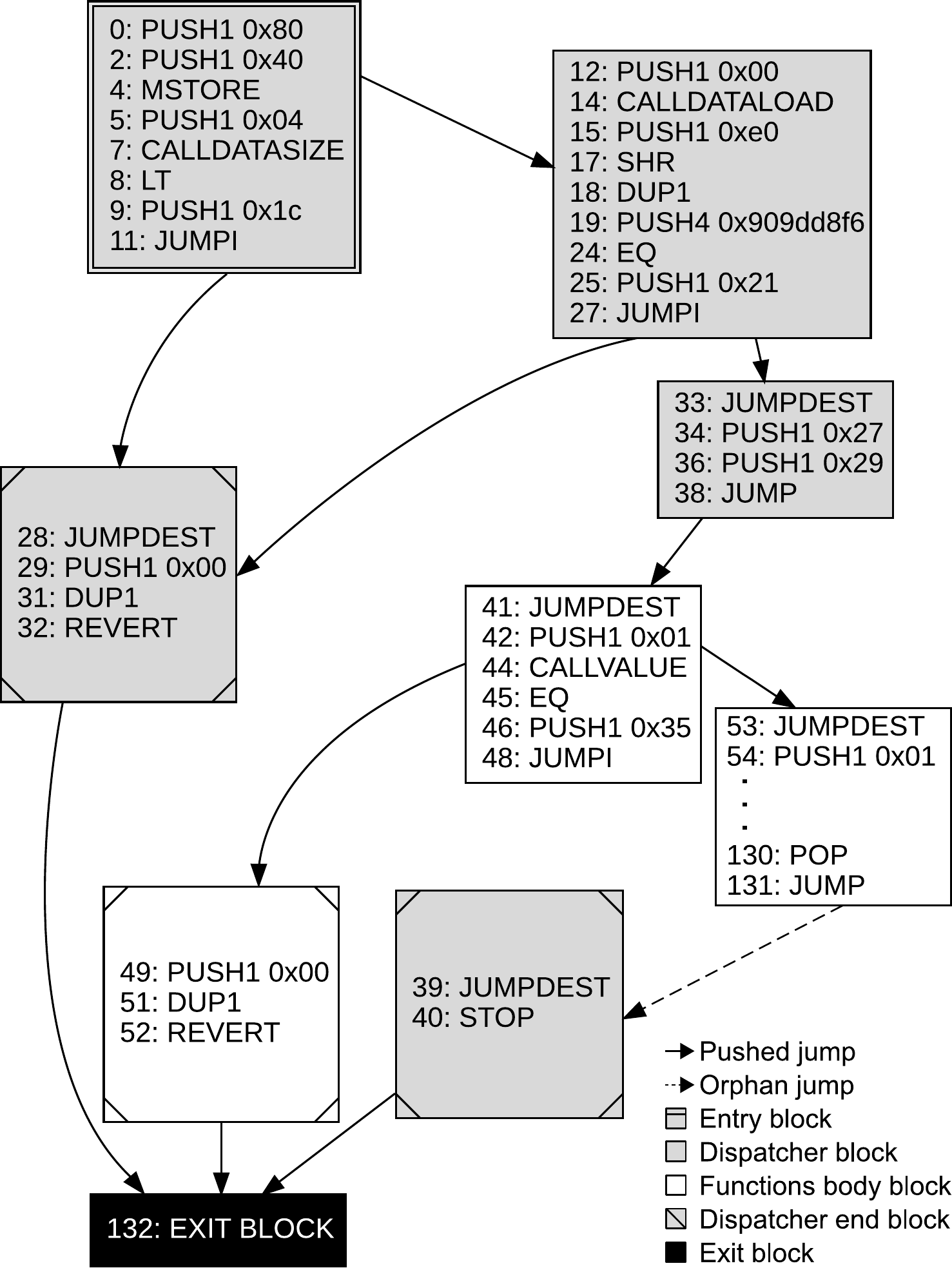}}
    \caption{Example of control-flow graph.}
    \label{fig:example-cfg}
\end{figure}

\subsection{Bytecode Parsing}

The analysis starts with the raw bytecode. 
The metadata section is identified by finding the respective header reported in the official documentation~\cite{soliditydoc}. 
In case of metadata with {\em experimental features}, the header is different and not documented. We inferred the header structure of non documented experimental cases by manually inspecting the bytecode of some contracts. The version of the Solidity compiler used to compile the bytecode is extracted from the metadata.

The metadata are then dropped and the remaining bytes are considered as actual code and parsed by the \ethersolve{} parser module. 
An example of how bytecode is parsed into opcodes is shown in Listing~\ref{lst:bytecode} and Listing~\ref{lst:opcodes}, where every two characters of the bytecode are translated into the respective opcode (e.g. \texttt{0x6080} becomes \texttt{PUSH1 0x80} and so on). 
Each opcode is univocally identified by its offset address.

\subsection{Basic Block Identification and Pushed Jumps}

A basic block is a sequence of opcodes which are executed consecutively between a jump target and a jump instruction, without any other instruction that alters the flow of control. 
Thus, opcodes that alter the control flow of the program divide the code into basic blocks. Opcodes \texttt{JUMP}, \texttt{JUMPI}, \texttt{STOP}, \texttt{REVERT}, \texttt{RETURN}, \texttt{INVALID}, \texttt{SELFDESTRUCT} mark the end of a basic block, whereas \texttt{JUMPDEST} marks the beginning of a new basic block. Every basic block is uniquely identified by its offset, i.e., the position of its first opcode in the bytecode. In Fig.~\ref{fig:example-cfg} we can see the basic blocks of the code in  Listing~\ref{lst:opcodes} extracted following this procedure. Indeed, each basic block either starts with a \texttt{JUMPDEST} or ends with an opcode which alters the control flow.

Once the code is divided into basic blocks, we proceed with the computation of the edges. This operation is not always simple as the jump destination is not an opcode parameter but it is available on top of the stack at execution time.
%
We identified two types of jumps: \emph{\pushed{} jumps} and \emph{orphan jumps}. A \emph{pushed jump} is a \texttt{JUMP} immediately preceded by a \texttt{PUSH} opcode, so that its target is easy to resolve, just by looking at the value in the preceding \texttt{PUSH} opcode. \emph{Orphan jumps} instead are not preceded by a \texttt{PUSH} and their target is not immediate to compute. In the example CFG of Fig.~\ref{fig:example-cfg} the block 53 ends with an \emph{orphan jump} whereas the remaining jumps are \emph{pushed jumps}. 

We start by computing the edges of \emph{\pushed{} jumps}.
To this end, each basic block is analysed according to its last opcode:
\begin{itemize}
    \item \texttt{JUMP} preceded by a \texttt{PUSH}: the argument of the push is the destination offset of the jump and the corresponding edge is added to the CFG. 
    \item \texttt{JUMPI} preceded by a \texttt{PUSH}: the false branch goes to the following block (in offset order), the true branch is the argument of the push interpreted as destination offset for the \texttt{JUMPI}. In this case the two corresponding edges are added to the CFG.
    \item \texttt{JUMP} not immediately preceded by a \texttt{PUSH}: the resolution of the jump is not trivial and it needs to be resolved through symbolic stack execution, described in \ref{ssec:symbolic-execution}.
    \item \texttt{REVERT}, \texttt{SELFDESTRUCT}, \texttt{RETURN}, \texttt{INVALID}, \texttt{STOP}: there are no successors as the control flow is interrupted.
    \item In any other case the control flow proceeds with the basic block in the next position in the bytecode.
\end{itemize}
At the end of this phase we have extracted a partial CFG where the edges related to \emph{orphan jumps} are still unresolved. For example, the extraction of the CFG of the code in Listing \ref{lst:solidity} at this point is depicted by the basic blocks and continuous edges of the CFG in Fig.~\ref{fig:example-cfg}, while the outgoing edge from the basic block 53 has not been resolved yet.

\subsection{Symbolic Stack Execution and Orphan Jumps}\label{ssec:symbolic-execution}

The most challenging step in the CFG construction is the resolution of the destinations of \emph{orphan jumps}. These jumps are very common: for instance the Solidity compiler uses them on return from function call. Indeed, between the function entry point and the function exit point (i.e., the return), the stack is heavily used by the function body to implement all the desired features (arithmetic operations, calls to other functions, transfer of funds).

The analysis consists in executing the stack symbolically: the algorithm walks the partially built CFG executing only the opcodes that interact with the jump addresses, updating the state of the stack accordingly, in such a way that the \emph{orphan jump} destinations can be found on the symbolic stack. Indeed, the symbolic stack execution considers only the opcodes in the \texttt{PUSH}, \texttt{DUP} and \texttt{SWAP} families, together with the \texttt{AND} and \texttt{POP} opcodes. For every other opcodes the symbolic stack pops and pushes ``unknown'' elements, as they do not deal with the jump addresses.

\begin{figure}[btp]
    \centerline{\includegraphics[width=\columnwidth]{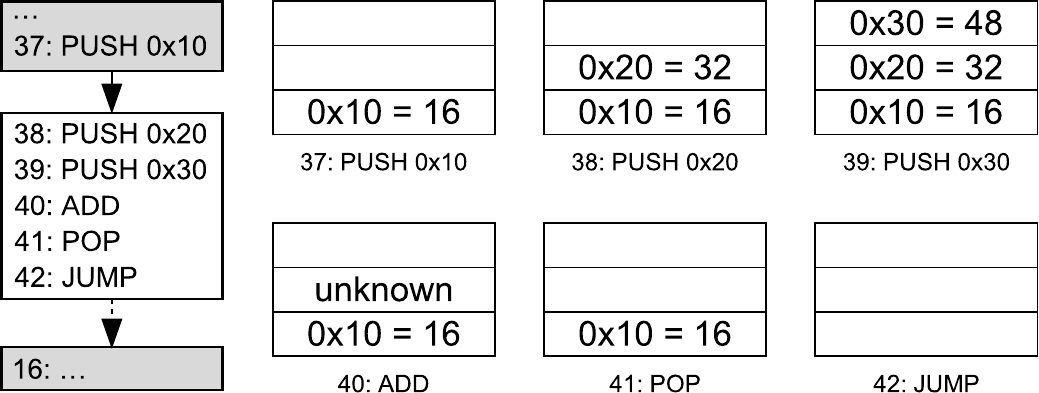}}
    \caption{Symbolic stack execution}
    \label{fig:example-stack-simple}
\end{figure}

In the example shown in Fig.~\ref{fig:example-stack-simple} there is a simple piece of code that has been executed symbolically to highlight the procedure. In particular the \texttt{ADD} is not modelled in full detail, but it simply consumes two elements and then generates a single ``unknown'' value. The jump address instead is loaded before the arithmetic operations, but it persists until the actual \texttt{JUMP} opcode, so it can be resolved.

The symbolic stack execution handles the opcodes according to the rules represented in the look-up Table~\ref{tab:symbolic-stack}, where the following notation is used:
\begin{itemize}
    \item $S$: stack that can contain numeric values or ``unknown'' to represent unknown values. We indicate the top of the stack as position $0$.
    \item $\alpha: Opcodes \rightarrow \mathbb{N}$: function that associates to each opcode the number of elements added to the stack.
    \item $\delta: Opcodes \rightarrow \mathbb{N}$: function that associates to each opcode the number of elements consumed from the stack.
\end{itemize}

\begin{table}
    \centering
    \captionof{table}{Look-up table for \textit{executeOpcode} in Algorithm~\ref{alg:orphanJumps}}
    \label{tab:symbolic-stack}
    \begin{tabular}{c c c c c}
    \toprule
        \textbf{Value} & \textbf{Name} & \textbf{$\delta$} & \textbf{$\alpha$} & \textbf{Stack} \\
    \toprule
        0x16 & AND & 2 & 1 & $S[1] = S[1] \land S[0]; S.pop()$ \\
    \midrule
        0x50 & POP & 1 & 0 & $S.pop()$ \\
    \midrule
        0x60 & PUSH1 & & & \\
        \vdots & PUSH$n$ & 0 & 1 & $S.push(opcode.\text{argument})$\\
         & & & & $|opcode.\text{argument}| = n \; \text{bytes}$ \\
        0x7f & PUSH32 & & & \\
    \midrule
        0x80 & DUP1 & & & \\
        \vdots & DUP$n$& $n$ & $n+1$ & $S.push(S[n-1])$\\
         & & & & \\
        0x0f & DUP16 & & & \\
    \midrule
        0x90 & SWAP1 & & & \\
        \vdots & SWAP$n$ & $n+1$ & $n+1$ & $S[0],S[n] = S[n], S[0]$ \\
         & & & & \\
        0x9f & SWAP16 & & & \\
    \midrule
        0x* & $O$ & $\delta(O)$ & $\alpha(O)$ & $S.pop() \enspace \delta(O) \; \text{times}$ \\
        & & & & $S.push(\text{{\scriptsize unknown}}) \enspace \alpha(O) \; \text{times}$\\
    \bottomrule
    \end{tabular}
\end{table}

The algorithm walks through the CFG using a DFS (Depth-First Search) keeping a state of the stack for each basic block. The following constraints have been enforced in order to avoid infinite loops: an edge cannot be analysed more than once with the same symbolic stack state, and there is a limit on the number of elements to compare when checking for stack equivalence. 

Another important aspect is the fact that a function can be called in different points in the code, resulting in different symbolic stacks and different paths on the CFG. In order to avoid infeasible paths (paths that real executions would never traverse) when the DFS visit encounters a basic block ending with a \texttt{JUMP} only its destination block, obtained by the symbolic stack, is added to the DFS queue.

The detailed algorithm for solving orphan jumps is shown in Algorithm~\ref{alg:orphanJumps}. It starts at Line 2 by initialising variable $V$ that stores the edges that have already been analysed using stack equivalence as described before (lines 20 and 27); an edge is also labelled with the symbolic stack that has been used for its symbolic execution (Cf. lines 19 and 26). 
Then, the queue $Q$ used for the DFS is initialised at Line 5: it contains pairs with a block and a symbolic stack. The first pair ${\left<CB, S\right>}$ contains the first block and an empty stack. Then, the algorithm proceeds with the symbolic stack execution by iteratively repeating the following steps until $Q$ is empty: 

\begin{itemize}
    \item Lines 9-11: symbolically execute the opcodes of the basic block and update the state of the symbolic stack according to the look-up Table~\ref{tab:symbolic-stack}.
    \item Lines 12-16: resolution of the \textit{orphan jump} destination with the newly updated symbolic stack. The target block is added as a successor of the basic block under analysis.
    \item Lines 17-31: handle the update of the queue $Q$. If the edge from the analysed basic block to the target one has not been already analysed with the same stack then the successor blocks are added to $Q$. If the last opcode is a \texttt{JUMP} then only its target block is added to $Q$.
\end{itemize}

\begin{algorithm}
\caption{Resolve Orphan Jumps}
\small 
\label{alg:orphanJumps}
\begin{algorithmic}[1]
    \Function{resolveOrphanJumps}{$basicBlocks$}
        \State $V \gets Set()$ \Comment{Visited}
        \State $CB \gets basicBlocks.first$ \Comment{Current block}
        \State $S \gets SymbolicExecutionStack()$ \Comment{Stack}
        \State $Q \gets Stack()$ \Comment{DFS queue}
        \State $Q.push({\left<CB, S\right>})$ \Comment{DFS first element}
        
        \While {$Q \neq \emptyset$}
            \State $CB, S \gets Q.pop()$
            
            \For {$op \in{} CB.opcodes$}
                \State $S.executeOpcode(op)$ \Comment{with look-up table} \label{alg:orphanJump:executeOpcode}
            \EndFor
            
            \If {$CB.opcodes.last == \texttt{JUMP}$}
                \State $NO \gets S.peek{}$
                \Comment{get next offset from stack}
                \State $NB = basicBlocks[NO]$ \Comment{Next block}
                \State $CB.addSuccessor(NB)$
            \EndIf
            
            \If {$CB.opcodes.last \neq \texttt{JUMP}$}
                \For {$suc \in CB.successors$}
                    \State $edge \gets{} \left< CB.\,\textit{offset}, suc.\,\textit{offset}, S \right>$
                    \If {$edge \notin V$}
                        \State $V.add(edge)$
                        \State $Q.push{\left<suc, S\right>}$
                    \EndIf
                \EndFor
            \ElsIf {$CB.opcodes.last == \texttt{JUMP}$}
                \State $edge \gets{} \left<CB.\,\textit{offset}, NO, S\right>$
                \If {$edge \notin V$}
                    \State $V.add(edge)$
                    \State $Q.push(\left<NB, S\right>)$
                \EndIf
            \EndIf
        \EndWhile
    \EndFunction
\end{algorithmic}
\end{algorithm}

\begin{figure}[btp]
    \centerline{\includegraphics[width=\columnwidth]{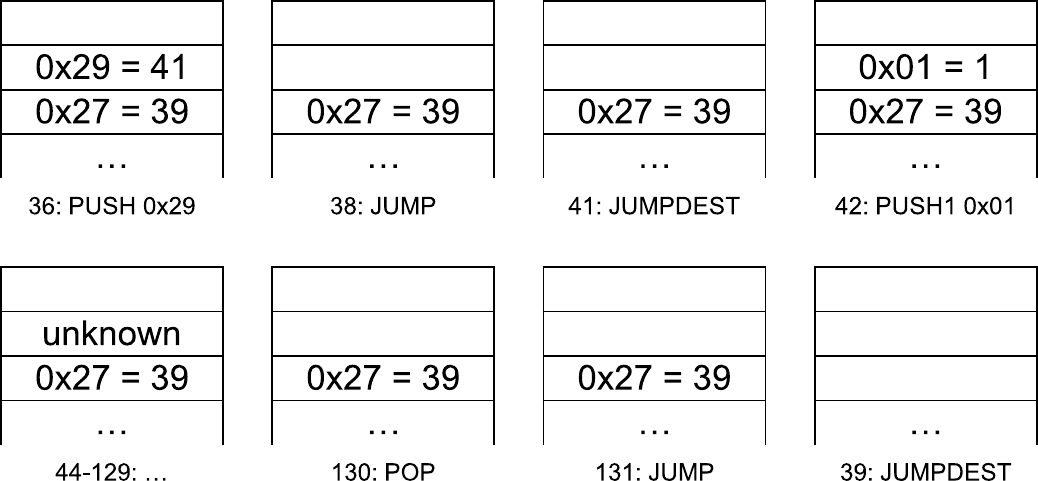}}
    \caption{Symbolic stack execution for orphan jump resolution}
    \label{fig:example-stack-running}
\end{figure}

An example of the symbolic stack execution for the resolution of \emph{orphan jumps} is shown in Fig.~\ref{fig:example-stack-running} and refers to a portion of the program shown in Listing \ref{lst:opcodes}, whose CFG is depicted in Fig.~\ref{fig:example-cfg}. The symbolic execution starts at the offset 36, which loads into the stack the value \texttt{0x29} after the value \texttt{0x27}. Then, our approach symbolically executes the \texttt{JUMP} opcode that, according to Table~\ref{tab:symbolic-stack}, consumes a value. Next, the symbolic execution of \texttt{JUMPDEST} leaves the stack unchanged and then value \texttt{0x01} is loaded. Then, the execution proceeds until the opcode at 129 following the look-up table, which leaves an unknown value on the stack that is removed by the \texttt{POP}. Finally, the opcode at 131 contains the \emph{orphan jump}, which can be resolved with the value pushed into the stack back on offset 34. At this point the symbolic stack execution can detect that the successor basic block is the number 39.

Eventually, we have resolved the target of all branches in the CFG, so the dashed edge in Fig. \ref{fig:example-cfg} is added at the end of this phase.

\subsection{Static Data Separation}
The proposed approach proceeds with the removal of static data (if present).
\ethersolve{} searches for the first basic block containing the instruction \texttt{0xFE}, which is the designated opcode for an invalid instruction. In fact, the Solidity compiler uses this opcode to mark the end of the executable code section and the beginning of the static data section. Thus, all the opcodes from here on are removed from the graph, and considered as static data and not as code. Then, the algorithm proceeds by removing from the graph any basic block that is not connected to the main graph, if any.

The static data detected are usually strings contained in the source code or child contracts which are instantiated by the main one through the opcodes \texttt{CREATE} and \texttt{CALL}. Even if in principle we cannot be sure that the removed data is  actually data and not code, our experimental validation shows that our approximation is very accurate.

\subsection{CFG Decoration}

In order to attach more information to the CFG that could be relevant to an analyst or for a subsequent static analysis (e.g., for vulnerability detection), \ethersolve{} tries to highlight some relevant portions, such as the dispatcher, the fallback function and the last basic block.

The dispatcher is the entry point of the smart contract, so it is at the beginning of the bytecode. The dispatcher directs the execution to the intended Solidity function and it manages parameters and return values. 
The fact that the dispatcher manages the return values is the key used for its detection. In fact, the only basic blocks that contain instructions such as \texttt{RETURN} and \texttt{STOP} are part of the dispatcher. This opcodes cannot be present in other locations as they would manage return values outside the dispatcher. Hence, the algorithm considers as dispatcher every basic block with an address lower then the address of these opcodes. In the example of Fig.~\ref{fig:example-cfg}, the dispatcher blocks are highlighted in grey.

This approach is effective to identify both the {\em linear} dispatcher, used in the older versions of the Solidity compiler, and the {\em tree} dispatcher, introduced in the latest versions of Solidity in order to improve the dispatcher performance.

The detection of the fallback function entry point is more difficult, because its structure has been changing continuously across different versions of the Solidity compiler. 
The first check of the dispatcher is the presence of call data and, if missing, it moves the execution to the Fallback function. Hence, the currently implemented technique starts from the entry block searching for the highest offset successor; then its successor with the highest offset is considered fallback only if it does not end with a \texttt{REVERT}, otherwise that would mean that the fallback function has not been declared or has been declared with only the \texttt{revert()} statement. This approach however does not work with some versions of \textit{solc} \todonew{due to different compilation patterns.}


The last step of the CFG decoration is the addition of an artificial unique exit point, for all the basic blocks with no successor. This could be useful for any static analysis that applies afterwards. This particular basic block in the example in Fig.~\ref{fig:example-cfg} is the number 132.




\todonew{\subsection{Re-entrancy Detection} \label{ssect:re-entrancy-detection}}

\todonew{Ethersolve has been pipelined with a subsequent static analysis meant to detect cases of the re-entrancy vulnerability. The proposed approach is a scan of the CFG in order to detect potential flows of execution where a
\texttt{SSTORE} opcode (which updates the contract state) is executed after a \texttt{CALL} opcode. This pattern is considered unsafe if the contract address where funds are transferred to by the {\tt CALL} cannot be statically determined by the symbolic stack execution. Indeed, in this case the funds destination could be controlled by an attacker who mounts an attack to exploit a re-entrancy vulnerability.

}




\section{Experimental Validation}
\label{sec:experiment}

In this section we present the results of our empirical validation of the accuracy of the CFG computed by \ethersolve{}. 
Three research questions guide the definition of our experimental validation:
\begin{itemize}
\item {\bf RQ$_1$}: What is the success rate of \ethersolve{} when analysing real\todonew{-}world smart contracts compiled with different versions of the Solidity compiler?
\item {\bf RQ$_2$}: How does \ethersolve{} compare with existing approaches?
\item \todonew{{\bf RQ$_3$}: How precise is the re-entrancy vulnerability detection built on top of \ethersolve{}?}
\end{itemize}
The first research question investigates the extent to which \ethersolve{} can process instances of real smart contracts with no errors. We are interested in verifying this on a wide range of smart contracts directly taken from the blockchain. The second one compares our approach with the state of the art. \todonew{The third research question compares the results of the re-entrancy vulnerability detection based on \ethersolve{} with other existing vulnerability detection tools.}

\subsection{Dataset}
\label{ssec:CaseStudies}
The empirical validation has been conducted using a dataset of 
smart contracts obtained from the list of verified contracts from Etherscan\footnote{Contracts have been download on 2020-06-10~\cite{etherscan-contracts}}. They are publicly available open source smart contracts with information about compilation, deployment and transactions. From this list we have randomly extracted 1000 contracts\footnote{Using the standard random library in Python 3.7}. Using Etherscan APIs, both bytecode and relevant information have been downloaded, obtaining for each smart contract its name, address, hash, deployment date, bytecode length, compiler version and other information that are not relevant for us. We enforced that smart contract bytecode hashes are unique, so that in the dataset there are no duplicates. Indeed, it is common practice to reuse existing smart contracts, especially libraries and interfaces, and deploy them multiple times in the blockchain at different accounts.

\begin{figure}[btp]
    \centering
    \includegraphics[width=\columnwidth]{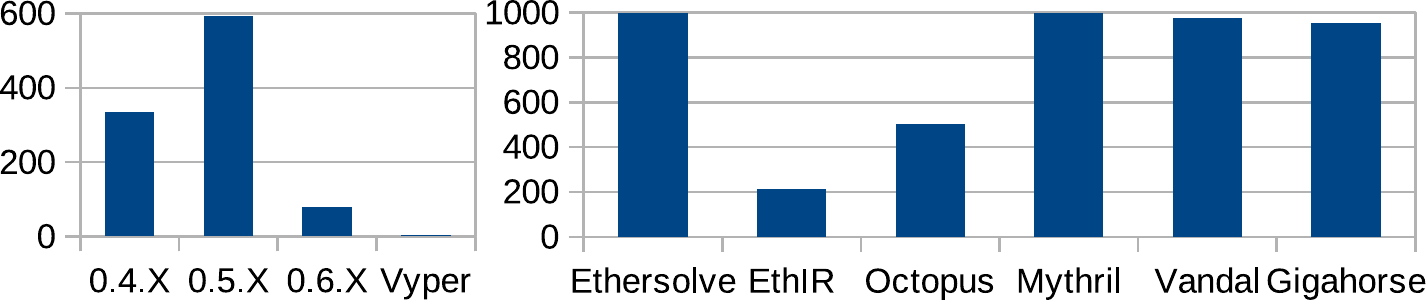}
    \caption{Compiler versions of dataset contracts (a) and success rate for the different tools (b)}
    \label{fig:versions-successrate}
\end{figure}

The average length of these smart contracts is \numprint{7351} bytes, the average transaction number is 337, the average balance is $9.6 \times 10^{17}$ Wei, which corresponds to 1158\$ (exchange rate on 2021-01-17). 
As shown in Fig.~\ref{fig:versions-successrate}.a, the compiler versions varies on a wide range, focusing especially on the old versions. This datum is crucial in order to assess that \ethersolve{} does not assume a specific Solidity version (the compiler often underwent dramatic changes) but our patterns are general and work well across many different language versions.

\subsection{Comparison Tools}

Among the existing tools for smart contracts analysis, we select for the comparison with \ethersolve{} the ones that: (i) perform a static analysis at the bytecode level (no information from the source code or ABI file), (ii) emit the CFG as output, possibly using a (documented) specific running configuration (e.g., a special command line flag).
These criteria led us to consider the following approaches:
\begin{itemize}
    \item \textit{Oyente-EthIR}: EthIR extends the Oyente framework and performs a high-level analysis of Ethereum bytecode. Oyente builds a CFG of Ethereum bytecode to detect different kinds of vulnerabilities~\cite{ethir},\cite{ethir-github}. 

    \item \textit{Octopus}: analysis framework for Ethereum bytecode. It starts from the bytecode and produces a CFG to support reverse engineering and understand the internal behaviour of smart contracts~\cite{octopus-github}.

    \item \textit{Mythril}: security analysis tool for Ethereum bytecode that detects security problems in smart contracts. It does not build a CFG, but a trace tree given by symbolic execution and SMT (Satisfiability Modulo Theories) solving~\cite{mythril},\cite{mythril-github},\cite{smt-paper}.

    \item \textit{Vandal}: static analysis framework for Ethereum smart contracts that decompiles bytecode to an intermediate representation that includes the control flow of the code~\cite{vandal2},\cite{vandal},\cite{vandal-github}.
    
    \item \todonew{\textit{Gigahorse}: a decompiler that transforms smart contracts from EVM bytecode into a highlevel 3-address code representation. The tool does not require the Solidity source code \cite{gigahorse1},\cite{madmax1},\cite{madmax2}.}
\end{itemize}

We discarded other tools such as \textit{Securify2}, \textit{Ethersplay}, \textit{Manticore} and \textit{Slither} because they analyse Solidity source code instead of Ethereum bytecode; \textit{Evm\_cfg\_builder} because we did not find an easy way to make it emit the CFG; \textit{Jeb} and \textit{MythX} because they are paid tools and \textit{Porosity} because it requires the ABI as input and also because the project seems discontinued. \todonew{\textit{Panoramix}~\cite{panoramix} was discarded because it decompiles the code without building a CFG and it is a discontinued project.}

\subsection{RQ$_1$: Success Rate} 

\ethersolve{} manages to analyse all the smart contracts in the dataset except three, obtaining a success rate of $99.7 \%$. \todonew{The success rate is defined as the ratio between the smart contracts analysed without critical errors and the size of the dataset.} The reason for these failures is that these three smart contracts do not match the most common patterns generated by the Solidity compiler. Indeed, one of them was written in Vyper rather than in Solidity, another one was empty with length of zero bytes and the last one has a unusual begin section that the tool failed to parse. The compiler version of Solidity is correctly identified by \ethersolve{} for all the solved contracts.

We measured the time spent by \ethersolve{} to parse the bytecode, generate the basic blocks, solve \emph{orphan jumps} and decorate the CFG. The average time is about 3 seconds per smart contract. For 930 smart contracts the analysis took less than a second each, while there are 7 smart contracts that needed more than a minute of computation, with a maximum of 10 minutes, due to their huge dimension and the large number of edges.

\subsection{RQ$_2$: Comparison with Existing Approaches}

First of all we compared the approaches in terms of success rate. We ran the tools on our dataset and we counted the successful executions (no crash) and the non-empty CFGs given in output. We also insert a reasonable timeout of 10 minutes per contract.


As shown in Fig.~\ref{fig:versions-successrate}.b and in Table~\ref{tab:automatic-analysis}, Mythril was able to analyse almost all the smart contracts (997 out of 1000), immediately followed by Vandal with 978 smart contracts \todonew{and by Gigahorse with 951 smart contracts.} Octopus and EthIR instead reached an error state on many smart contracts, computing a CFG only for 504 and 212 smart contracts respectively. 
Next we focused on the CFGs emitted by the tools, starting with an automatic numerical comparison of nodes and edges. Then, we proceeded with a manual inspection of the anomalous cases.

For each smart contract in the dataset we adopt a common representation for the outputs of the different tools. The chosen representation is a JSON file that contains a list of nodes (which represent the basic blocks), identified by their offset and a list of edges, identified by a pair of offsets.
For each smart contract we count the number of nodes, the number of edges and the differences in the numbers of nodes and edges among the CFGs generated by different tools. To understand if there are portions of the bytecode which are actually data (static data, child contracts or compiler metadata) but that the candidate tool wrongly interprets as code, we calculate the number of nodes in the CFG generated by a candidate tool that have a higher offset than the highest offset node obtained by \ethersolve{}.
The candidate tools have been executed in Docker~\cite{docker} containers for compatibility reasons.

In the manual analysis we focused on the smart contracts for which the automatic analysis reported uncommon or anomalous results. These are relatively few contracts with sensitive differences in the number of nodes or edges between the CFG extracted by \ethersolve{} and by the compared tool. To support the manual analysis we implemented a script that generates a \emph{diff graph}, where the two CFGs are combined and the nodes that are present only in the first or in the second graph are highlighted in a different colour.

The results of both automatic and manual analysis are discussed in the following. Table~\ref{tab:automatic-analysis} contains a summary of the automatic analysis, with the smart contracts successfully analysed, the average number of both nodes and edges and the average number of basic blocks in the static data segment. Table~\ref{tab:manual-analysis} contains a subset of interesting/anomalous contracts that have been subject to manual analysis. Since contract names are not unique, contracts are identified by their addresses.

\begin{table}[btp]
    \centering
    \captionof{table}{\todonew{Comparison with state of the art.}}
    \label{tab:automatic-analysis}
    \begin{tabular}{l | c c c c}
         \toprule
            & \textbf{Success Rate} & \textbf{Average} & \textbf{Avg. Nodes} & \textbf{Average} \\
            & \textbf{(/1000)} & \textbf{Nodes} & \textbf{Static data} & \textbf{Edges} \\
         \midrule
            EthIR & 212 & 139.4 & 51.6 & 150.3 \\
            Octopus & 504 & 241.4 & 11.8 & 220.7 \\
            Mythril & 997 & 4.0 & 8.5 & 3.1 \\
            Vandal & 978 & 302.6 & 6.7 & 493.6 \\
            Gigahorse & 951 & 245.4 & 1.2 & 288.7 \\
            {\bf \ethersolve{}} & {\bf 997} & {\bf 301.6} & {\bf 0} & {\bf 361.8} \\
        \bottomrule
    \end{tabular}
\end{table}

\begin{table}[btp]
    \captionof{table}{\todonew{Subset of smart contracts used in manual analysis}}
    \label{tab:manual-analysis}
    \begin{tabular}{l l}
        \toprule
            \multicolumn{2}{c}{\textbf{Octopus}}                                       \\
        \midrule
            a. SafeMath            & 0x7de33b2672efb11fde366dae96bd63b985bce186 \\
            b. ZipmexTokenP.      & 0xaa602de53347579f86b996d2add74bb6f79462b2 \\
            c. ltdFDTfactory     & 0xf155152d838b7a023317ad8c1e8c02aab7e8f2a2 \\
            d. Dividend           & 0xdb6f50cf0c521a98b6852839aa5cbea4e2430052 \\
        \midrule
            \multicolumn{2}{c}{\textbf{Mythril}}                                       \\
        \midrule
            e. DharmaKeyR.     & 0x0000000000bda2152794ac8c76b2dc86cba57cad \\
            f. DharmaT.R.      & 0x8b028e2fad2dc99999fb784ca9d7267981c90b4d \\
            g. AltesF.G. & 0x8313675d1405f3f7aee3da9d63e0bf5c30c75832 \\
            h. SMRT16Ext         & 0xdabb0c3f9a190b6fe4df6cb412ba66c3dd3e2ad1 \\
        \midrule
            \multicolumn{2}{c}{\textbf{Vandal}}                                        \\
        \midrule
            i. FlashFloss         & 0xfd4085e56a96787fb7acd9b49510f874c3d4afcb \\
            j. FoxInvSplit         & 0xd69015163e250a70ee4a607812afda5372132cc4 \\
            k. BCN20              & 0x1964f2f3ce45ac518b18ef4aa4265f8aadcef4ae \\
        \midrule
            \multicolumn{2}{c}{\textbf{Gigahorse}}                                        \\
        \midrule
            l. OneSplit              & 0x3a2d9db352580eb50018fc86eae32e19070a9982\\
            m. EthexLoto         & 0x0e26b2dc8ef577baf50891eac94f0def59b5da16 \\
            n. ManagedAccount         & 0x0f4f45f2edba03d4590bd27cf4fd62e91a2a2d6a\\
            o. ERC20Salary              & 0xcc2ba2eac448d60e0f943ebe378f409cb7d1b58a\\
        \bottomrule
    \end{tabular}
\end{table}

\subsubsection{Oyente-EthIR}
In most cases, when EthIR finds more nodes than \ethersolve{}, they have very high offset (Cf. Table~\ref{tab:automatic-analysis}), so they are static data or metadata interpreted as code. Instead, when nodes match, edges match too. Because of its very low success rate (21\%), we deemed not so interesting to continue with a manual analysis of this tool results.

\subsubsection{Octopus}
Like EthIR, Octopus finds more blocks than \ethersolve{} with a higher offset, so they are probably data and not code. In the big majority of the cases, however, Octopus finds very few edges, sometimes even zero edges (Cf. Table~\ref{tab:automatic-analysis}). During manual analysis, we discovered that Octopus misses some patterns for the metadata separation, so metadata are parsed as if it were code (Table~\ref{tab:manual-analysis}.a). Moreover, in many cases, Octopus does not detect edges that should be present according to source code (Table~\ref{tab:manual-analysis}.b). In some other contracts, Octopus evaluates the bytecode as creation code, thus it analyses only the second part starting with \texttt{60806040} (the most common begin sequence of a contract); however, this part of the code is a child contract and not the main one, so the computed graphs are completely different (Table~\ref{tab:manual-analysis}.c and Table~\ref{tab:manual-analysis}.d).

\subsubsection{Mythril}
This tool is a particular case, as it does not extract a CFG, but a trace tree from dynamic/symbolic analysis in order to detect vulnerabilities. The output of this execution trace is not directly comparable with the CFG computed by \ethersolve{}. Nonetheless, indirect comparison can be performed by checking if the \ethersolve{} CFG misses nodes or edges that are found by the Mythril dynamic analysis. This case would correspond to incompletenesses in the CFG elaborated by \ethersolve{}.

In the 4\% of the contracts, Mythril finds a bunch of edges which are not detected by \ethersolve{}. In some cases, Mythril adds some artificial basic blocks containing \texttt{"0: STOP"} which are placeholders that do not come from the analysed code, but they indicate the end of the execution trace (Table~\ref{tab:manual-analysis}.e). Sometimes, basic blocks are not split when there is a \texttt{JUMPDEST} in the middle, so there is a little discrepancy in the edges, but the CFG is definitely compatible (Table~\ref{tab:manual-analysis}.f and Table~\ref{tab:manual-analysis}.g). In some other cases, Mythril finds new basic blocks and edges that are not part of the main contract, but that are opcodes of a child contract created by the main one (Table~\ref{tab:manual-analysis}.h). The child contract code is computed at runtime, so the \ethersolve{} static analysis simply treats those bytes as part of the static data segment.

\subsubsection{Vandal}
Vandal CFG has always one node more than \ethersolve{}, which is the ending node with the \texttt{INVALID} opcode. In the 36\% of the solved contracts the edges match, but in the remaining cases there are differences that we analysed manually.
In some contracts the basic blocks do not match because Vandal does not support two opcodes that have been added in the most recent versions of the EVM~\cite{selfbalance-eip}. In fact, the \texttt{SELFBALANCE} opcode is treated as invalid by Vandal, obtaining a basic block break with no outgoing edges (Table~\ref{tab:manual-analysis}.i and Table~\ref{tab:manual-analysis}.j). In many other cases, Vandal detects a huge amount of edges. Probably, when Vandal is not able to correctly compute the destination of a jump, it conservatively assumes all the basic blocks as possible successors (Table~\ref{tab:manual-analysis}.k). This hypothesis is supported by the diff graph, which shows too many outgoing edges for basic blocks that occur at the end of functions, probably because the return address could not be computed accurately. However, the number of call sites (that should correspond to the number of return edges) is much smaller according to the Solidity source code of these contracts. All in all, as shown in Table~\ref{tab:automatic-analysis}, the average number of edges found by Vandal is significantly higher than \ethersolve{}, whereas the basic blocks are reconstructed in a similar way.

\subsubsection{\todonew{Gigahorse}}
\todonew{Gigahorse records a good success rate, it is able to correctly analyse 95.1\% of the samples. 
It tries to identify the private functions inside the code, and the computed CFG reflects this objective. Because of this strategy, the conversion into the intermediate representation is tricky; often its CFGs contain artificial blocks with the special \texttt{CALLPRIVATE} statement, introduced by Gigahorse to mark private function calls. Nevertheless, automatic comparison records high similarity with \ethersolve{}, so we proceeded with the deeper manual inspection.

For the contract shown in Table~\ref{tab:manual-analysis}.l, \ethersolve{} computes a set of basic blocks that are unreachable from the contract entry point, that might represent dead code. However, these unreachable blocks are not present in the Gigahorse CFG. Similar cases are Table~\ref{tab:manual-analysis}.m and Table~\ref{tab:manual-analysis}.n, where \ethersolve{} finds more nodes and edges than Gigahorse; these elements correspond to a portion in the middle of the bytecode which has not been reported in Gigahorse's CFGs.
An opposite case happens on the contract in Table~\ref{tab:manual-analysis}.o, where \ethersolve{} identifies a set of blocks that are not reachable from the contract entry point whereas, according to the Gigahorse CFG, these blocks are reachable. Our speculation is that such basic blocks (which have a higher offset) belong to a child contract or to an internal library which is not a proper part of the main contract (e.g. called via \texttt{STATICCALL}), and thus they are skipped by \ethersolve{} (that only analyses intra-contract calls).}


\todonew{\subsection{RQ$_3$: Re-entrancy Vulnerability Detection}}
\todonew{In order to validate the effectiveness of the vulnerability detector built on top of \ethersolve{}, we compared it with the benchmark shared by Ghaleb and Pattabiraman~\cite{ghaleb2020effective}.
The benchmark consists in 50 Ethereum smart contracts whose source code has been injected with re-entrancy vulnerabilities coming from 42 code snippets. This benchmark has been used by Ghaleb and Pattabiraman to compare the most prominent vulnerability detection tools.
While their comparison was performed at source-code level, \ethersolve{} targets the bytecode level, so these injected contracts have been compiled before applying our analysis.
While the original dataset consisted of 50 files, each source file could contain more than one contract and injected vulnerabilities could multiply in the compiled contracts, because of the use of inheritance that caused vulnerable code to be cloned from abstract contracts to concrete ones. Additionally, abstract contracts do not produce bytecode as they are not executable.
Hence, to simplify the analysis, we opted to scan only one compiled contract per source file, i.e.\ the largest compiled contract, assuming that the remaining contracts were only supporting libraries or abstract contracts, whose CFGs were disconnected from the main one.


Another problematic aspect, acknowledged by Ghaleb and Pattabiraman, is that the dataset already contained vulnerabilities before SolidiFI injection, but they were not documented. To gather comparable results, we considered only vulnerabilities added by SolidiFI injection, by running \ethersolve{} before and after injection, and by keeping only those new vulnerabilities that are detected by the second scan and not by the first scan.} Table~\ref{tab:re-entrancy-comparison} shows the results of the comparison, measuring the average number of detected vulnerabilities for each sample by each tool and then computing the difference with the average number of bugs injected by SolidiFI (\texttildelow26.86).
The table highlights that \ethersolve{} is the second-best analysis tool, immediately after Slither. After a deeper comparison with the results given by these two top tools, we discovered that \ethersolve{} misses some cases (due to the choice of analysing only a single bytecode) whereas Slither often finds false positives. Indeed, the samples where \ethersolve{} found false negatives are smart contract obtained by multi-contract source code. Moreover, \ethersolve{} is the only tool that analyses bytecode, thus having less information, so its results are even more valuable: \ethersolve{} can successfully analyse even closed-source smart contracts with high precision.


\begin{table}[btp]
    \centering
    \captionof{table}{\todonew{Summary of re-entrancy analysis comparison for the different tools~\cite{ghaleb2020effective} and for \ethersolve{}}}
    \label{tab:re-entrancy-comparison}
    \begin{tabular}{l|r|r}
        \toprule
        Tool & Avg. detection per sample & Diff. with SolidiFI injection \\
        \midrule
        Manticore & 2.14 & -24.72 \\
        Oyente & 8.58 & -18.28 \\
        Mythril & 15.12 & -11.74 \\
        Slither & 27.26 & +0.40 \\
        Securify & 28.84 & +1.98 \\
        \textbf{\ethersolve} & \textbf{26.00} & \textbf{-0.86} \\
        \bottomrule
    \end{tabular}
\end{table}

\todonew{To confirm the precision of \ethersolve{}, we scanned all the 42 code snippets that SolidiFI uses for the injection. The \ethersolve{} analysis exactly detected all the 42 vulnerabilities, with a precision of 100\%.}

\subsection{Discussion}

The results obtained in this experimental validation suggest that \ethersolve{} is very effective in computing an accurate CFG: it is able to work on a wide range of Solidity versions and in almost all cases it computes an exhaustive graph.

The key point of our approach is the simplicity of the symbolic stack execution, which is limited to only a tiny set of opcodes, but which is capable of resolving the destinations of \emph{orphan jumps}. However, there are particular cases of very complex or big smart contracts with peculiar structures for which \ethersolve{} is not able to identify certain edges.

\todonew{Among the compared tools, only Gigahorse showed an accuracy similar to \ethersolve{}. However, they seems to be complementary, because each one could solve cases that the other one could not.}




\todonew{The results of the vulnerability detector suggest that \ethersolve{} is a powerful tool, and that can be easily extended to support accurate subsequent static analyses based on a precise CFG. Indeed, the performance of a simple reachability analysis compares to those of state-of-the-art security scanning tools.}
\section{Related Work}
\label{sec:related}

Over the last five years many tools have been developed to analyse Ethereum smart contracts, with different approaches and different objectives. A recent survey of them has been written by Praitheeshan et al.~\cite{security-analysis-methods}.

Some tools analyse directly the bytecode, often trying to build a CFG. Their approaches are very similar, as they try to execute symbolically the code to create logic predicates which, once resolved with a solver such as \emph{Z3 theorem prover}~\cite{z3-github}, can determine the destinations of the \emph{orphan jumps}. Among these tools there are \emph{Oyente}~\cite{oyente}, \emph{EthIR}~\cite{ethir} and \emph{Octopus}~\cite{octopus-github}. However, these tools aim at detecting vulnerabilities, and the extracted CFG is only an intermediate output. 

A slightly different approach is the one used by \emph{Vandal}~\cite{vandal, vandal2}, which translates the bytecode into registry based operations, identifies the basic blocks and then tries to resolve the jump address through a fixed point analysis. Even in this case the CFG is only an intermediate output, as the target of Vandal is the vulnerability analysis with the \emph{Souffle} suite~\cite{souffle}.
Our tool instead focuses on the CFG building, keeping the symbolic stack execution as simple as possible, in order to resolve the highest number of \emph{orphan jumps}.

A related tool which extract a CFG is \textit{Jeb}~\cite{jeb}, which is a professional decompiler with the ability to analyse Ethereum smart contracts. However, it is closed source with a subscription fee. Another decompiler is \textit{Porosity}, one of the first tool to analyse Ethereum bytecode, but it needed the contract ABI too. Moreover it is discontinued since January 2018~\cite{porosity}. 
\todonew{A relevant decompiler is Gigahorse~\cite{gigahorse1, madmax1, madmax2}, a recent tool which builds a CFG and tries to find internal function with heuristics, obtaining an approximation of the original Solidity source code. Another decompiler is Panoramix~\cite{panoramix} which, however, does not emit a CFG.}

The wide majority of the Ethereum tools that performs vulnerability analysis do not expose a CFG, or even they do not extract it. Other tools instead do the analysis on the Solidity source code, or use the bytecode together with additional information that are not always available for closed source contracts.
A completely different approach is the one implemented by \textit{Mytril}, which uses symbolic execution, SMT solving and taint analysis to detect a variety of security vulnerabilities~\cite{mythril-github}. It does not build a CFG, but a trace tree, i.e. a representation of all the execution paths encountered during the analysis. Its objective is to detect as many vulnerabilities as possible.
\textit{Crytic}~\cite{crytic} is an application that collects many tools for smart contract analysis, such as \textit{Manticore}, \textit{Ethersplay}, \textit{Echidna}, \textit{Slither} and more, but they do not use a CFG or they do not analyse only the bytecode. In fact their objective is the vulnerabilities detection inside the Solidity source code. 

Finally, there are other tools such as \textit{Securify2}~\cite{securify2}, which analyse Solidity source code, \textit{Maian}~\cite{maian}, which performs dynamic analysis on a private blockchain, and \textit{Gasper}~\cite{gasper}, which analyse the gas cost of the contracts.
\section{Conclusion}
\label{sec:conclusion}


Despite it would be very important to automatically analyse smart contracts and detect potential defects and vulnerabilities on their code, most of the existing analysis tools for Ethereum bytecode come with some shortcomings and limitations. For example, the accurate extraction of the CFG from the Ethereum bytecode is very challenging due to engineering decision on its infrastructure.
%
We propose \ethersolve{}, a fast, reliable and precise approach to compute an accurate CFG from Ethereum bytecode. We believe that this CFG could be the starting point for new static analysis tools that aim at detecting defects and vulnerabilities in Ethereum smart contracts, built on top of an accurate CFG, \todonew{such as the \ethersolve{} re-entrancy detector.} 

\section*{Acknowledgment}
This paper has been partially supported by project MIUR 2018-2022 “Dipartimenti di Eccellenza”.


\bibliographystyle{IEEEtran}
\bibliography{main}

\end{document}